\documentclass[conference]{IEEEtran}
\IEEEoverridecommandlockouts
\usepackage{cite}
\usepackage{amsmath,amssymb,amsfonts}
\usepackage{algorithmic}
\usepackage{graphicx}
\usepackage{textcomp}
\usepackage{xcolor}

\DeclareMathAlphabet{\mathsfbr}{OT1}{cmss}{m}{n}
\SetMathAlphabet{\mathsfbr}{bold}{OT1}{cmss}{bx}{n}
\DeclareRobustCommand{\msf}[1]{%
  \ifcat\noexpand#1\relax\msfgreek{#1}\else\mathsfbr{#1}\fi
}

\makeatletter
\newcommand{\msfgreek}[1]{\csname s\expandafter\@gobble\string#1\endcsname}
\makeatother

\DeclareFontEncoding{LGR}{}{} 
\DeclareSymbolFont{sfgreek}{LGR}{cmss}{m}{n}
\SetSymbolFont{sfgreek}{bold}{LGR}{cmss}{bx}{n}
\DeclareMathSymbol{\salpha}{\mathord}{sfgreek}{`a}
\DeclareMathSymbol{\sbeta}{\mathord}{sfgreek}{`b}
\DeclareMathSymbol{\sgamma}{\mathord}{sfgreek}{`g}
\DeclareMathSymbol{\sdelta}{\mathord}{sfgreek}{`d}
\DeclareMathSymbol{\sepsilon}{\mathord}{sfgreek}{`e}
\DeclareMathSymbol{\szeta}{\mathord}{sfgreek}{`z}
\DeclareMathSymbol{\seta}{\mathord}{sfgreek}{`h}
\DeclareMathSymbol{\stheta}{\mathord}{sfgreek}{`j}
\DeclareMathSymbol{\siota}{\mathord}{sfgreek}{`i}
\DeclareMathSymbol{\skappa}{\mathord}{sfgreek}{`k}
\DeclareMathSymbol{\slambda}{\mathord}{sfgreek}{`l}
\DeclareMathSymbol{\smu}{\mathord}{sfgreek}{`m}
\DeclareMathSymbol{\snu}{\mathord}{sfgreek}{`n}
\DeclareMathSymbol{\sxi}{\mathord}{sfgreek}{`x}
\DeclareMathSymbol{\somicron}{\mathord}{sfgreek}{`o}
\DeclareMathSymbol{\spi}{\mathord}{sfgreek}{`p}
\DeclareMathSymbol{\srho}{\mathord}{sfgreek}{`r}
\DeclareMathSymbol{\ssigma}{\mathord}{sfgreek}{`s}
\DeclareMathSymbol{\stau}{\mathord}{sfgreek}{`t}
\DeclareMathSymbol{\supsilon}{\mathord}{sfgreek}{`u}
\DeclareMathSymbol{\sphi}{\mathord}{sfgreek}{`f}
\DeclareMathSymbol{\schi}{\mathord}{sfgreek}{`q}
\DeclareMathSymbol{\spsi}{\mathord}{sfgreek}{`y}
\DeclareMathSymbol{\somega}{\mathord}{sfgreek}{`w}

\DeclareMathSymbol{\svarsigma}{\mathord}{sfgreek}{`c}

\DeclareMathSymbol{\sGamma}{\mathalpha}{sfgreek}{`G}
\DeclareMathSymbol{\sDelta}{\mathalpha}{sfgreek}{`D}
\DeclareMathSymbol{\sTheta}{\mathalpha}{sfgreek}{`J}
\DeclareMathSymbol{\sLambda}{\mathalpha}{sfgreek}{`L}
\DeclareMathSymbol{\sXi}{\mathalpha}{sfgreek}{`X}
\DeclareMathSymbol{\sPi}{\mathalpha}{sfgreek}{`P}
\DeclareMathSymbol{\sSigma}{\mathalpha}{sfgreek}{`S}
\DeclareMathSymbol{\sUpsilon}{\mathalpha}{sfgreek}{`U}
\DeclareMathSymbol{\sPhi}{\mathalpha}{sfgreek}{`F}
\DeclareMathSymbol{\sPsi}{\mathalpha}{sfgreek}{`Y}
\DeclareMathSymbol{\sOmega}{\mathalpha}{sfgreek}{`W}

\DeclareRobustCommand{\mcal}[1]{%
  \ifcat\noexpand#1\relax\mathnormal{#1}\else\cal{#1}\fi
}
\DeclareRobustCommand{\BM}[1]{%
  \ifcat\noexpand#1\relax\bm{\boldUppercaseItalicGreek{#1}}\else\bm{#1}\fi
}
\makeatletter
\newcommand{\boldUppercaseItalicGreek}[1]{\csname var\expandafter\@gobble\string#1\endcsname}
\makeatother

\newcommand{\V}[1]{\bm{#1}} 
\newcommand{\Set}[1]{{\mcal{#1}}} 

\usepackage{bm}
\usepackage{color, courier}
\usepackage{psfrag}
\usepackage{acronym}
\usepackage{amsmath}
\usepackage{amssymb}
\usepackage{amsfonts}
\usepackage[colorlinks = true, allcolors = black]{hyperref}
\usepackage{latexsym}
\usepackage{mathrsfs}
\usepackage{epstopdf}
\usepackage{algorithm}
\usepackage{mathtools}
\usepackage[caption=false, font=scriptsize]{subfig}
\usepackage{algorithmic}
\usepackage{relsize}
\usepackage{comment}
\usepackage[left = .57in, right = .57in, top = .7in, bottom = .95in]{geometry}

\graphicspath{{figures/}}

\DeclareMathOperator*{\argmin}{arg\,min}

\newcommand{\st}{\operatorname{s.t.}\,}

\newtheorem{theorem}{Theorem}

\definecolor{green}{rgb}{0, 0.5, 0}
\definecolor{pink}{rgb}{1, 0, 1}

\definecolor{bg}{RGB}{199, 237, 204}

\acrodef{agi}[AgI]{augmented information}
\acrodef{ldp}[LDP]{Lyapunov drift-plus-penalty}
\acrodef{lp}[LP]{linear programming}
\acrodef{ap}[AP]{access point}
\acrodef{pdf}[pdf]{probability density function}
\acrodef{ue}[UE]{user equipment}
\acrodef{sfc}[SFC]{service function chain}
\acrodef{alg}[ALG]{augmented layered graph}
\acrodef{ER}[STAR]{star route}
\acrodef{mec}[MEC]{multi-access edge computing} 
\acrodef{csi}[CSI]{channel state information}

\acrodef{nfv}[NFV]{network function virtualization}
\acrodef{sdn}[SDN]{software defined networking}

\acrodef{wrt}[w.r.t.]{with respect to}
\acrodef{wlog}[w.l.o.g.]{Without loss of generality}

\acrodef{umw}[UMW]{universal max-weight}

\def\BibTeX{{\rm B\kern-.05em{\sc i\kern-.025em b}\kern-.08em
    T\kern-.1667em\lower.7ex\hbox{E}\kern-.125emX}}

\begin{document}

\title{
    Dynamic Control of Data-Intensive Services over Edge Computing Networks
}

\author{
    \IEEEauthorblockN{
        Yang Cai\IEEEauthorrefmark{1},
        Jaime Llorca\IEEEauthorrefmark{2},
        Antonia M. Tulino\IEEEauthorrefmark{2}\IEEEauthorrefmark{3},
        Andreas F. Molisch\IEEEauthorrefmark{1}
    } 
    \IEEEauthorblockA{
        \IEEEauthorrefmark{1}University of Southern California, CA 90089, USA. Email: \{yangcai, molisch\}@usc.edu
    }
    \IEEEauthorblockA{
        \IEEEauthorrefmark{2}New York University, NY 10012, USA. Email: \{jllorca, atulino\}@nyu.edu
    }
    \IEEEauthorblockA{
        \IEEEauthorrefmark{3}University\`{a} degli Studi di Napoli Federico II, Naples 80138, Italy. Email: antoniamaria.tulino@unina.it
    }
}

\maketitle

\IEEEpubid{
\begin{minipage}{3\columnwidth}
    \centering
    {\footnotesize
    \vspace{50pt}
    This work has been submitted to the IEEE for possible publication. Copyright may be transferred without notice, after which this version may no longer be accessible.
    }
\end{minipage}
}

\begin{abstract}
Next-generation distributed computing networks (e.g., edge and fog computing) enable the efficient delivery of {\em delay-sensitive}, {\em compute-intensive} applications by facilitating access to computation resources in close proximity to end users. Many of these applications (e.g., augmented/virtual reality) are also {\em data-intensive}: in addition to user-specific ({\em live}) data streams, they require access to ({\em static}) digital objects (e.g., image database) to complete the required processing tasks. When required objects are not available at the servers hosting the associated service functions, they must be fetched from other edge locations, incurring additional communication cost and latency. In such settings, overall service delivery performance shall benefit from jointly optimized decisions around (i) routing paths and processing locations for live data streams, together with (ii) cache selection and distribution paths for associated digital objects. In this paper, we address the problem of dynamic control of data-intensive services over edge cloud networks. We characterize the network stability region and design the first throughput-optimal control policy that coordinates processing and routing decisions for both live and static data-streams. Numerical results demonstrate the superior performance (e.g., throughput, delay, and resource consumption) obtained via the novel multi-pipeline flow control mechanism of the proposed policy, compared with state-of-the-art algorithms that lack integrated stream processing and data distribution control.
\end{abstract}

\section{Introduction}

The class of \ac{agi} services refers to a wide range of services and applications designed to deliver information of real-time relevance that results from the online aggregation, processing, and distribution of multiple data streams. \ac{agi} services such as system automation (e.g., smart homes/factories/cities, self-driving cars) and Metaverse experiences (e.g., multiplayer gaming, immersive video, virtual/augmented reality) are driving unprecedented requirements for communication, computation, and storage resources \cite{cai2022metaverse}. To address this need, distributed cloud network architectures such as \ac{mec} are becoming a promising paradigm, providing end users with efficient access to nearby computation resources. Together with continued advances in network virtualization and programmability \cite{SDN_NFV13}, distributed cloud networks allow flexible and elastic deployment of disaggregated services composed of multiple software functions that can be dynamically instantiated at distributed network locations.

The associated problem of service function chain (SFC) orchestration has 
received significant attention in the recent literature. One main line of work studies this problem in a static setting,  where the goal is to allocate multi-dimensional (i.e., communication, computation, storage) network resources for function/data placement and flow routing, in order to optimize a network-wide objective, e.g., maximizing the number of accepted requests \cite{huang2021throughput} or minimizing the operational cost \cite{yue2021resource}. While useful for long timescale end-to-end service optimization, the corresponding formulations typically take the form of (NP-Hard) mixed integer programs that lead to algorithms with either high complexity or sub-optimal performance, and overlook the dynamic nature of service demands and network conditions.

To address the problem in a dynamic scenario, \cite{FenLloTulMol:J18a} introduced the {\em dynamic cloud network control} problem, where one needs to make online packet processing, routing, and scheduling decisions in response to stochastic system states (e.g., service demands and resource capacities). Among existing techniques, Lyapunov-drift control \cite{Nee:B10} is a widely-used approach to design throughput-optimal algorithms \cite{cai2020mec,cai2021delay,cai2021multicast,cai2022delay_arxiv,cai2022multicast_arxiv}, e.g., DCNC \cite{FenLloTulMol:J18a} and UCNC \cite{zhang2021multicast}, by {\em dynamically} exploring routing and processing diversity. In general, centralized routing policies, e.g., UCNC, that exploit global knowledge to guarantee packets follow {\em acyclic} paths, can attain better delay performance than distributed counterparts, e.g., DCNC. We refer the reader to a longer version of this paper \cite{cai2022CCC_arxiv} for a more comprehensive literature review.

An increasingly relevant feature of next-generation \ac{agi} services is their 
{\em intensive data requirements}: in addition to {\em live data} streams, task (or service function) processing also requires access to pre-stored {\em static data} objects. For example, in an augmented reality (AR) application, access to pre-stored scene objects is required to generate augmented/enhanced images or video streams (see Fig. \ref{fig:service}). While the aforementioned studies on dynamic cloud network control optimize the live data stream (routing and processing) pipeline without considering the access to static data, other works in data-centric computing networks \cite{yeh2021deco} have addressed the static data distribution and processing problem without considering the live data pipeline.

To close the gap, this paper addresses the problem of ``online delivery of {\em data-intensive \ac{agi} services} over edge computing networks'', which require real-time aggregation, processing, and distribution of live and static data streams. We design a control policy that {\em jointly} decides (i) routing paths and processing locations for live streams, and (ii) cache selection and distribution paths for associated static objects, to optimize end-to-end service delivery performance. The main challenges come from the coupling of live and static data routing with the associated function processing decisions. To the best of our knowledge, this is the first paper that studies joint processing and routing control of multiple pipelines for the online delivery of stream-processing services.

The contributions of this work are summarized as follows:
\begin{itemize}
    \item We characterize the stability region of distributed cloud networks supporting data-intensive \ac{agi} service delivery.
    \item We design the first throughput-optimal control policy for online data-intensive service delivery, termed DI-DCNC, which coordinates processing and routing decisions for live and static data streams.
    \item We conduct illustrative numerical experiments to demonstrate the superior performance of DI-DCNC.
\end{itemize}

\section{System Model}

\subsection{Cache-Enabled MEC Network Model}

Consider a distributed cloud network, modeled by a directed graph $\Set{G} = (\Set{V}, \Set{E})$, with $\Set{V}$ and $\Set{E}$ denoting the node and edge sets, respectively. Each vertex $i\in \Set{V}$ represents a node equipped with computation resources (e.g., edge server) for service function processing. Each edge $(i, j) \in \Set{E}$ represents a point-to-point communication link, which can support data transmission from node $i$ to $j$. Let $\delta^-(i)$ and $\delta^+(i)$ denote the incoming and outgoing neighbor sets of node $i$, respectively.

Time is slotted, and the network processing and transmission capacities are quantified as follows. (i) $C_i$: the maximum number of processing instructions (e.g., floating point operations) that can be executed in one time slot at node $i$. (ii) $C_{ij}$: the maximum number of data units (e.g., packets) that can be transmitted in one time slot over link $(i, j)$.

We assume that the network nodes are also equipped with storage resources to cache databases composed of digital objects whose access may be required for service function processing. In this paper, we focus on cache selection and routing decisions with fixed database placement and refer the reader to \cite{cai2022CCC_arxiv} for extensions that include database placement/replacement policies. Let $\Set{V}(k) \subset \Set{V}$ denote the {\em static sources} of database $k$, i.e., the set of nodes that cache database $k$.

\subsection{Data-Intensive Service Model}
\label{sec:dag_model}

In this paper, we illustrate the proposed algorithm in the context of a single-function data-intensive service. The generalization to multiple service functions is briefly described in Section \ref{sec:extensions} and presented in detail in \cite{cai2022CCC_arxiv}.

\begin{figure}[t]
    \centering
    \includegraphics[width = .98\columnwidth]{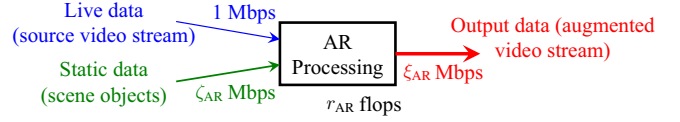}
    \caption{
    Example AR application with one processing function that takes two inputs (live and static data) to create augmented data.
    }
    \vspace{-10pt}
    \label{fig:service}
\end{figure}

Each service $\phi$ includes one task (or service function) that must process {\bf coupled} user-specific data and associated digital objects -- referred to as {\em live packets} and {\em static packets}, respectively -- for the generation of consumable streams. As shown in Fig. \ref{fig:service}, an example AR application is composed of an AR processing function that must process the user live data (source video stream) together with associated static objects (scene objects) to create the output augmented data (augmented video stream). 

The function of service $\phi$ is described by $4$ parameters $(\xi_\phi, r_\phi, k_\phi, \zeta_\phi)$, defined as:
(i) {\em object name} $k_\phi$: the name (or index) of the database to which the static object belongs,
(ii) {\em merging ratio} $\zeta_\phi$: the number of static packets per input live packet,
(iii) {\em workload} $r_{\phi}$: the amount of computation resource (e.g., instructions per time slot) to process one input live packet, and
(iv) {\em scaling factor} $\xi_\phi$: the number of output packets per input live packet.

\subsection{Client Model}
\label{sec:packet_management}

We define each client $c$ by a $3$-tuple $(s, d, \phi)$, denoting the source node $s$ (where the live packets arrive to the network), the destination node $d$ (where the output packets are requested for consumption), and the requested service $\phi$, respectively.

\subsubsection{Live Packet Arrival}

Let $a^{(c)}(t)$ be the number of live data packets of client $c$ arriving to the network at time $t$. For each client $c$, we assume the arrival process $\{a^{(c)}(t): t\geq 0\}$ is i.i.d. over time, with mean arrival rate of $\lambda^{(c)}$, and bounded maximum arrival number.

\subsubsection{Static Packet Provisioning}

Upon a live packet arrival, one static source $v \in \Set{V}(k_{\phi})$ must be selected to produce a copy of the required static packet.

We refer to a live packet and its associated static packet required for its processing as belonging to the same {\em packet-level request}.

\subsection{Queuing System}

Each packet (live or static) arriving to the network gets associated with a route for its delivery, and we establish actual queues to accommodate packets waiting for processing or routing.
For each link $(i, j) \in \Set{E}$, we create one {\em transmission} queue collecting all packets waiting to cross the link. In addition, a novel {\bf paired-packet queuing system} is constructed at each node $i \in \Set{V}$, composed of:
(i) the {\em processing} queue collecting the {\em paired} live and static packets concurrently present at node $i$, which are ready for joint processing, and
(ii) the {\em waiting} queue collecting the {\em unpaired} live or static packets waiting for their in-transit associates, which are not qualified for processing until joining the processing queue upon their associates' arrivals.

\section{Policy Space and Network Stability Region}

\subsection{Transformation to an Augmented Layered Graph}

We propose an \ac{alg} model extending the layered graph \cite{zhang2021multicast}, to analyze and optimize the data-intensive \ac{agi} service delivery problem.

\subsubsection{Topology}

The \ac{alg} is composed of three pipelines, referred to as {\em live}, {\em static} and {\em output} pipelines, and each pipeline has the same topology as the actual network $\Set{G}$, except for the static pipeline that includes an additional node $o_1'$ and its outgoing edges to all static sources.
We note that:
(i) The live, static, and output pipelines accommodate exogenously arriving live packets, in-network replicated static packets, and processed output packets, respectively, and represent their associated routing over the network.
(ii) In the static pipeline, we create a {\em super static source} node $o_1'$ connected to all static sources $v \in \Set{V}(k_{\phi})$, which is equivalent to assume that {\em $o_1'$ is the only static source of database $k_{\phi}$}.%
\footnote{
    If node $o_1'$ can provide static packets to $i_1'$ along a path $(o_1', v_1', \cdots, i_1')$ in the ALG; then, in the actual network, we can select $v$ to produce the packets and send them to node $i$ along the rest of the path. And vice versa.
}
(iii) The three pipelines are placed in two layers, describing packets {\em before} (layer $1$, henceforth indicated by subscript $1$) and {\em after} processing (layer $2$), respectively.
There are inter-layer edges connecting corresponding nodes, which represent processing operations, i.e., the live and static packets pushed through these edges are processed into the output packet in the actual network.

\begin{figure}[t]
    \centering
    \includegraphics[width = .9 \columnwidth]{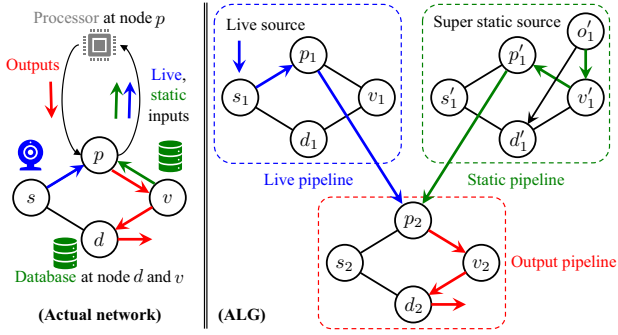}
    \caption{Illustration of the ALG model 
    for the delivery of AR service.}
    \vspace{-10pt}
    \label{fig:ALG}
\end{figure}

In the example in Fig. \ref{fig:ALG}, the live packet and static packet (which is generated 
via replication at the static source $v$) are routed following the blue and green paths to node $p$, respectively. After getting processed at node $p$, the output packet is delivered along the red path to destination $d$.

Mathematically, given the actual network $\Set{G}$, the \ac{alg} of service $\phi$ is given by $\Set{G}_{\phi} = (\Set{V}^{(\phi)}, \Set{E}^{(\phi)})$, with $\Set{V}^{(\phi)} = \{ i_1, i_1', i_2 : i\in \Set{V} \} \cup \{ o_1' \}$ and $\Set{E}^{(\phi)} = \Set{E}^{(\phi)}_{\text{tr}} \cup \Set{E}^{(\phi)}_{\text{pr}} \cup \Set{E}^{(\phi)}_{\text{st}}$, in which $\Set{E}^{(\phi)}_{\text{tr}} = \{ (i_1, j_1), (i_1', j_1'), (i_2, j_2) : (i, j) \in \Set{E} \}$, $\Set{E}^{(\phi)}_{\text{pr}} = \{ (i_1, i_2), (i_1', i_2) : i\in \Set{V} \}$, and $\Set{E}^{(\phi)}_{\text{st}} = \{ (o_1', v_1') : v\in \Set{V}(k_{\phi}) \}.$

\subsubsection{Flow Constraints}

Denote by $f_{\imath \jmath} \geq 0$ the flow rate associated with edge $(\imath, \jmath) \in \Set{E}^{(\phi)}$, defined as the average rate of packets traversing the edge (in {\em packets per slot}). In particular,
for $\forall \, (i, j) \in \Set{E}$, $f_{i_1 j_1}$, $f_{i_1' j_1'}$, $f_{i_2 j_2}$ represent the transmission rates of the live, static, and output packets over link $(i, j)$;
for $\forall \, v \in \Set{V}(k_{\phi})$, $f_{o_1' v_1'}$ is the local replication rate of the static packets at static source $v$;
for $\forall \, i \in \Set{V}$, $f_{i_1 i_2}$, $f_{i_1' i_2}$ denote the processing rates of the live and static packets at node $i$, which generates the output packets at node $i$ at rate $\xi_{\phi} f_{i_1 i_2}$ by the definition of ``scaling factor $\xi_{\phi}$'' in Section \ref{sec:dag_model}.

The flow rates must satisfy the following constraints.

(i) Live and output flow conservation: for $\forall\, i\in \Set{V}$,
\begin{subequations} \label{eq:flow_conservation_1} \begin{align}
    & \hspace{5pt} \sum\nolimits_{j \in \delta^{+}(i)} f_{i_1 j_1} + f_{i_1 i_2} = \sum\nolimits_{j \in \delta^{-}(i)} f_{j_1 i_1}, \label{eq:flow_conservation_live_in} \\
    & \sum\nolimits_{j \in \delta^{+}(i)} f_{i_2 j_2} = \sum\nolimits_{j \in \delta^{-}(i)} f_{j_2 i_2} + \xi_{\phi} f_{i_1 i_2}, \label{eq:flow_conservation_live_out}
\end{align} \end{subequations}
i.e., the total outgoing rate of packets that are transmitted and processed equals the total incoming rate of packets that are received and generated by processing, for live \eqref{eq:flow_conservation_live_in} and output \eqref{eq:flow_conservation_live_out} streams.

(ii) Static flow conservation: for $\forall\,i\in \Set{V}$,
\begin{align} \label{eq:flow_conservation_2}
    \sum_{j\in \delta^+ (i)} f_{i_1' j_1'} + f_{i_1' i_2} = \sum_{j\in \delta^- (i)} f_{j_1' i_1'} + f_{o_1' i_1'} \V{1}_{\{i \in \Set{V}(k_{\phi})\}},
\end{align}
which is interpreted the same as (i), except for that static packets can be generate by replication rather than processing.

(iii) Data 
merging: for $\forall\, i\in \Set{V}$,
\begin{align} \label{eq:merging_ratio}
  f_{i_1' i_2} = \zeta_{\phi} f_{i_1 i_2} ,
\end{align}
i.e., the processing rates of static and live packets are associated by the ``merging ratio $\zeta_{\phi}$'' as defined in Section \ref{sec:dag_model}.

\subsection{Policy Space}
\label{sec:policies}

Next, we define the space of policies for data-intensive service delivery, encompassing joint packet processing, routing, and replication decisions.
In line with \cite{zhang2021multicast} and the increasingly adopted software defined networking (SDN) paradigm, we focus on centralized routing and distributed scheduling policies.
Upon a packet-level request and associated live packet arrival at its source, the policy selects (i) a routing path and a processing location for the live packet, (ii) cache selection (i.e., chosen static source to replicate the static packet) and distribution path for associated static packet, and (iii) a routing path for the resulting output packet. In addition, each node/link needs to schedule packets for processing/transmission at each time slot.

\subsubsection{Decision Variables}

A policy thus includes two actions.

\underline{\bf Route Selection:} For each packet-level request, choose a set of edges in the ALG and associated flow rates satisfying \eqref{eq:flow_conservation_1} -- \eqref{eq:merging_ratio}, based on which:
(i) the cache selection decision is specified by the replication rate $f_{o_1' v_1'}$,
(ii) the routing path for each (live or static) packet is given by the edges with non-zero rates in the corresponding pipeline, and 
(iii) the processing location selection is indicated by the processing rates $f_{i_1 i_2}$ and $f_{i_1' i_2}$, which guarantee that the live and static packets meet at the same node $i$ due to \eqref{eq:merging_ratio}.

\underline{\bf Packet Scheduling:} At each time slot, for each node $i\in \Set{V}$ and link $(i, j)\in \Set{E}$, schedule packets from the local processing queues (which hold {\em paired} packets) and transmission queues, for operation, and the incurred resource consumption shall not exceed the corresponding capacities $C_i$ and $C_{ij}$.

\subsubsection{Efficient Policy Space} \label{sec:efficient_space}

We define an efficient policy space in which the routing path of each packet is required to be {\em acyclic}, without compromising the achievable performance (e.g., throughput, delay, and resource consumption).

As illustrated in Fig. \ref{fig:ALG}, for each request, we choose a \ac{ER} $\sigma$ in the \ac{alg}, including an internal node $p_2$ and three leaf nodes $s_1$, $o_1'$ and $d_2$ connecting to it via the path $\sigma_1 \cup (p_1, p_2)$, $\sigma_1' \cup (p_1', p_2)$ and $\sigma_2$, where $p_2$ denotes the {\em processing location}, and $\sigma_1$, $\sigma_1'$ and $\sigma_2$ denote the acyclic {\em routing paths} of the live (from $s_1$ to $p_1$), static (from $o_1'$ to $p_1'$), and output (from $p_2$ to $d_2$) packets, respectively.

For each client $c$, the set of all \acp{ER}, denoted by $\Set{F}_c$, is {\em finite}, and the route selection decision can be expressed by
\begin{align} \label{eq:routing_decision_rep}
    A(t) = \big\{ a^{(c, \sigma)}(t): \sum\nolimits_{\sigma\in \Set{F}_c} a^{(c, \sigma)}(t) = a^{(c)}(t),\ \forall\, c, \sigma \big\}
\end{align}
where $a^{(c, \sigma)}(t)$ denotes the number of requests of client $c$ that arrive at time $t$ and get associated with $\sigma$ for delivery.

\subsection{Network Stability Region}

We define the {\em network stability region} $\Lambda$ as the set of arrival vectors $\V{\lambda} = \{ \lambda^{(c)}: \forall\,c \}$ under which there exists a policy to stabilize the actual queues. The stability region describes the network capability to support service requests, and is characterized by the following theorem.

\begin{theorem} \label{thm:nsr}
An arrival vector $\V{\lambda}$ is interior to the stability region $\Lambda$ if and only if for each client $ c$, there exist probability values $\big\{ \mathbb{P}_c(\sigma) : \sum_{\sigma \in \Set{F}_c} \mathbb{P}_c(\sigma) = 1 \big\}$, such that: for $\forall\, i, (i, j)$,
\begin{align}
\sum_{c, \sigma} \lambda^{(c)} \rho_{i}^{(c, \sigma)} \mathbb{P}_c(\sigma) \leq C_{i},\ 
\sum_{c, \sigma} \lambda^{(c)} \rho_{ij}^{(c, \sigma)} \mathbb{P}_c(\sigma) \leq C_{ij}
\end{align}
where $\rho_i^{(c, \sigma)}$ and $\rho_{ij}^{(c, \sigma)}$ are the processing and transmission resource loads imposed on node $i$ and link $(i, j)$ if a packet of client $c$ is delivered via \ac{ER} $\sigma$, given by:
\begin{subequations} \label{eq:load} \begin{align}
\rho_i^{(c, \sigma)} & = r_{\phi} \bm{1}_{ \{ (i_1,\, i_2)\in \sigma \} } \\
\rho_{ij}^{(c, \sigma)} & = \bm{1}_{ \{ (i_1,\, j_1 )\in \sigma \} } + \zeta_{\phi} \bm{1}_{ \{ (i_1',\, j_1' )\in \sigma \} } + \xi_{\phi} \bm{1}_{ \{ (i_2,\, j_2 )\in \sigma \} }. \label{eq:link_load}
\end{align} \end{subequations}
\end{theorem}

\begin{IEEEproof}
See \cite[Theorem 1]{cai2022CCC_arxiv}.
\end{IEEEproof}

In \eqref{eq:link_load} , the three terms denote the resource loads imposed on link $(i, j)$ if the live, static, and output packets traverse $(i_1, j_1)$, $(i_1', j_1')$, and $(i_2, j_2)$ in \ac{ER} $\sigma$, respectively.

\section{Proposed Algorithm}

Next, we present the proposed {\em data-intensive dynamic cloud network control} (DI-DCNC) policy. We first introduce a single-hop virtual system (in Section \ref{sec:virtual_system}) to derive packet routing decisions (in Section \ref{sec:virtual}). Then, we present the packet scheduling policy, and summarize the actions taken in the actual network (in Section \ref{sec:ENTO}).

\subsection{Virtual System}
\label{sec:virtual_system}

\subsubsection{Precedence Constraint}

In line with \cite{SinMod:J18,zhang2021multicast}, we create a virtual network, where the {\em precedence constraint}, which imposes a packet to be transmitted hop-by-hop along its route, is relaxed by allowing a packet upon route selection to be immediately inserted into the virtual queues associated with all links in the route.

We emphasize that the virtual system is only used for route selection and NOT relevant to packet scheduling.

\subsubsection{Virtual Queues}

We create virtual queues $\tilde{Q}_i(t)$ for $\forall\, i\in \Set{V}$ and $\tilde{Q}_{ij}(t)$ for $\forall\, (i, j) \in \Set{E}$, to represent the resource loads of the nodes and links in the virtual system, which are interpreted as the {\em anticipated} resource loads in the actual network.

The queuing dynamics are given by:
\begin{subequations} \label{eq:virtual_q}
\begin{align}
\tilde{Q}_i(t+1) & = \big[ \tilde{Q}_i(t) - C_i + \tilde{a}_i(t) \big]^+ \\
\tilde{Q}_{ij}(t+1) & = \big[ \tilde{Q}_{ij}(t) - C_{ij} + \tilde{a}_{ij}(t) \big]^+
\end{align}
\end{subequations}
where $C_i$ and $C_{ij}$ are interpreted as the amount of processing/transmission resources that are ``served'' at time $t$; $\tilde{a}_i(t)$ and $\tilde{a}_{ij}(t)$ are ``additional'' resource loads imposed on the nodes and links by newly arriving requests. Recall that each request gets associated with a route for delivery upon arrival, which immediately impacts the queuing states of all links in the route, and thus:
\begin{align}
\tilde{a}_i(t) = \sum_{c, \sigma} \rho_i^{(c, \sigma)} a^{(c, \sigma)}(t),\ 
\tilde{a}_{ij}(t) = \sum_{c, \sigma} \rho_{ij}^{(c, \sigma)} a^{(c, \sigma)}(t)
\end{align}
with $\rho_i^{(c, \sigma)}$ and $\rho_{ij}^{(c, \sigma)}$ given by \eqref{eq:load}.

\subsection{Optimal Virtual Network Decisions}
\label{sec:virtual}

\subsubsection{Lyapunov Drift Control}

Next, we leverage Lyapunov drift control theory to derive a policy that stabilizes the {\em normalized} virtual queues $\V{Q}(t) = \big\{ Q_i(t) \triangleq \tilde{Q}_i(t) / C_i : \forall\,i\in \Set{V} \big\} \cup \big\{ Q_{ij}(t) = \tilde{Q}_{ij}(t) / C_{ij} : \forall\,(i, j) \in \Set{E} \big\}$, which have equivalent stability properties as the virtual queues and can be interpreted as {\em queuing delay} in the virtual system.

Define Lyapunov function $L(t) \triangleq \| \V{Q}(t) \|^2 / 2$, and its drift $\Delta(\V{Q}(t)) \triangleq L(t+1) - L(t)$. Then we can derive the following upper bound of the drift $\Delta(\V{Q}(t))$ (see \cite[Appendix B.1]{cai2022CCC_arxiv}):
\begin{align}
\label{eq:ldp}
\Delta(\V{Q}(t)) \leq B - \| \V{Q}(t)\|_1 + 
\sum\nolimits_{c,\sigma} O^{(c, \sigma)}(t) \, a^{(c, \sigma)}(t)
\end{align}
where $B$ is a constant, and $O^{(c, \sigma)}(t)$  is referred to as the weight of STAR $\sigma$, given by:
\begin{align}  \label{eq:ALG_weight}
& \hspace{-8pt}  O^{(c, \sigma)}(t)
= \sum_{(i, j)} \big[ w_{i_1 j_1}^{(c)}(t) \, \bm{1}_{ \{ (i_1, j_1)\in \sigma \} } + w_{i_1' j_1'}^{(c)}(t) \, \bm{1}_{ \{ (i_1', j_1')\in \sigma \} } \nonumber \\
& \hspace{15pt}  + w_{i_2 j_2}^{(c)}(t) \, \bm{1}_{ \{ (i_2, j_2)\in \sigma \} } \big] + \sum_{i\in \Set{V}} w_{i_1 i_2}^{(c)}(t) \bm{1}_{ \{ (i_1, i_2)\in \sigma \} }
\end{align}
in which $w_{i_1 j_1}^{(c)}(t) = Q_{ij}(t) / C_{ij}$, $w_{i_1' j_1'}^{(c)}(t) = \zeta_{\phi} (Q_{ij}(t) / C_{ij})$, $w_{i_2 j_2}^{(c)}(t) = \xi_{\phi} (Q_{ij}(t) / C_{ij})$, $w_{i_1 i_2}^{(c)}(t) = r_{\phi}(Q_i(t) / C_i)$, and we define $w_{i_1' i_2}^{(c)}(t) = w_{o_1' v_1'}^{(c)}(t) = 0$ for $\forall\, i\in \Set{V}, v\in \Set{V}(k_\phi)$.

The proposed algorithm is designed to minimize \eqref{eq:ldp} over the route selection decision $A(t)$ given by \eqref{eq:routing_decision_rep}, or equivalently,
\begin{align} \label{eq:ldp_opt}
    \min_{A(t)}\ \sum_{c, \sigma} 
    O^{(c, \sigma)}(t)\, a^{(c, \sigma)}(t),\ 
    \st\ \eqref{eq:routing_decision_rep}.
\end{align}

\subsubsection{Route Selection}

Given the linear structure of \eqref{eq:ldp_opt}, the optimal route selection decision is given by:
\begin{align} \label{eq:routing_decision}
    a^{\star\, (c, \sigma)}(t) = a^{(c)}(t) \V{1}_{ \{ \sigma = \sigma^\star \} },\ 
    \sigma^\star \triangleq \argmin_{\sigma \in \Set{F}_c} \, O^{(c, \sigma)}(t),
\end{align}
i.e., all requests of client $c$ arriving at time $t$ are delivered by the {\em min-STAR} $\sigma^\star$, i.e., the \ac{ER} with the minimum weight. 

The remaining problem is to find the min-STAR $\sigma^\star$. To this end, we create a {\em weighted \ac{alg}} in which we assign the weight $w_{\imath \jmath}^{(c)}(t)$ defined in \eqref{eq:ALG_weight} to each edge $(\imath, \jmath)$ in the \ac{alg}, under which the weight of the STAR $\sigma$, $O^{(c, \sigma)}(t)$, equals the sum of individual edge weights, given by:
\begin{align} \begin{split} \label{eq:ER_weight_2}
    O^{(c, \sigma)}(t) & = | \sigma_1 | + | \sigma_1' | + w_{p_1 p_2}^{(c)}(t) + | \sigma_2 |
\end{split} \end{align}
where $|\sigma_1|$, $|\sigma_1'|$, and $|\sigma_2|$ denote the weights of the routing paths of the live, static, and output packets in the weighted ALG, respectively, and $p$ is the processing location.

Based on \eqref{eq:ER_weight_2}, we propose to find the min-STAR in two steps. First, fix the processing location $p$, and we can treat the optimization of path weights for the live, static, and output packets as {\em separate} problems. For example, optimizing $|\sigma_1|$ is equivalent to finding the shortest path from $s_1$ to $p_1$ in the weighted \ac{alg}, and similar for $|\sigma_1'|$ (from $o_1'$ to $p_1'$) and $|\sigma_2|$ (from $p_2$ to $d_2$). Therefore, the minimum weight of all STARs {\em selecting node $p\in \Set{V}$ as the processing location} is
\begin{align} \begin{split} \label{eq:weight_j}
    W_p = \operatorname{SPW}(s_1, p_1) + \operatorname{SPW}(o_1', p_1') & \\
    + w_{p_1 p_2}^{(c)}(t) & + \operatorname{SPW}(p_2, d_2)
\end{split} \end{align}
where $\operatorname{SPW}(\imath, \jmath)$ denotes the weight of the shortest path $\operatorname{SP}(\imath, \jmath)$ from node $\imath$ to $\jmath$ in the weighted \ac{alg} (the path of static packet is given by $\operatorname{SP}(o_1', p_1') = {(o_1', v_1'(p))} \cup \operatorname{SP}(v_1'(p), p_1')$, with $v(p)$ denoting the selected static source). Then, we select the processing location $p^\star\in \Set{V}$ to minimize the weight \eqref{eq:weight_j}.

The procedure is summarized as follows:

\underline{\bf Route Selection:}
At each time slot $t$, all requests of client $c$ get associated with the \ac{ER} specified as:
\begin{subequations} \label{eq:route_selection} \begin{align}
& \text{Processing:} \ p^\star = \argmin\nolimits_{p\in \Set{V}} \, W_p\text{ (given by \eqref{eq:weight_j})}, \\
& \text{Cache selection:} \ v^\star = v(p^\star), \\
& \text{Transmission:} \ \underbrace{\operatorname{SP}(s_1, p_1^\star)}_{\text{live packet}},\, \underbrace{\operatorname{SP}(v^\star_1{}', p^\star_1{}')}_{\text{static packet}},\, \underbrace{\operatorname{SP}(p_2^\star, d_2)}_{\text{output packet}}.
\end{align} \end{subequations}

\subsubsection{Complexity Analysis}

The above procedure requires to run Dijkstra's algorithm to calculate the shortest path distance from $s_1$, $o_1'$ and $d_2$ to the other nodes in the \ac{alg}, with complexity $\mathcal{O} (|\mathcal{E}| \log |\mathcal{V}|)$ \cite{kleinberg2006algorithm}.

\subsection{Optimal Actual Network Decisions} \label{sec:ENTO}

Next, we present the control decisions in the actual network. We adopt the route selection decisions made in the virtual network \eqref{eq:route_selection}, and the extended nearest-to-origin (ENTO) policy \cite{SinMod:J18} for packet scheduling, described as follows.

\underline{\bf Packet Scheduling:}
{\em At each time slot, for each node / link, give priority to the packets in the corresponding processing (holding paired packets) / transmission queues which have crossed the smallest number of edges in the \ac{alg}.}

To sum up, the proposed {\bf DI-DCNC} algorithm operates as follows. At each time slot: (i) assign the \ac{ER} \eqref{eq:route_selection} to all requests of client $c$ for delivery, (ii) at each node/link, schedule packets for processing/transmission by ENTO, and (iii) update the virtual queues by \eqref{eq:virtual_q}.

\subsection{Performance Analysis}

\begin{theorem}
For any arrival vector $\V{\lambda}$ within the stability region $\Lambda$, the actual queues are rate stable under DI-DCNC.
\end{theorem}

\begin{IEEEproof}
See \cite[Theorem 2]{cai2022CCC_arxiv}.
\end{IEEEproof}

\section{Extensions to General Service Model}
\label{sec:extensions}

In general, a service can perform (i) multi-step processing (i.e., more than one service function) on (ii) multiple source streams (i.e., more than one live and static input).

To handle such services, we can incorporate more (i) layers and (ii) pipelines, into the ALG. The (extended) DI-DCNC algorithm follows the same rule for route selection, i.e., finding the min-weight route to deliver each packet-level request, and a key step is to select a {\em sequence} of processing locations, which transforms the problem into multiple unicast problems. We refer the reader to \cite{cai2022CCC_arxiv} for detailed descriptions and derivations.

\section{Numerical Results}

\begin{figure*}
    \centering
    \begin{minipage}[t]{0.24\textwidth}
        \centering
        \includegraphics[width=.95\textwidth]{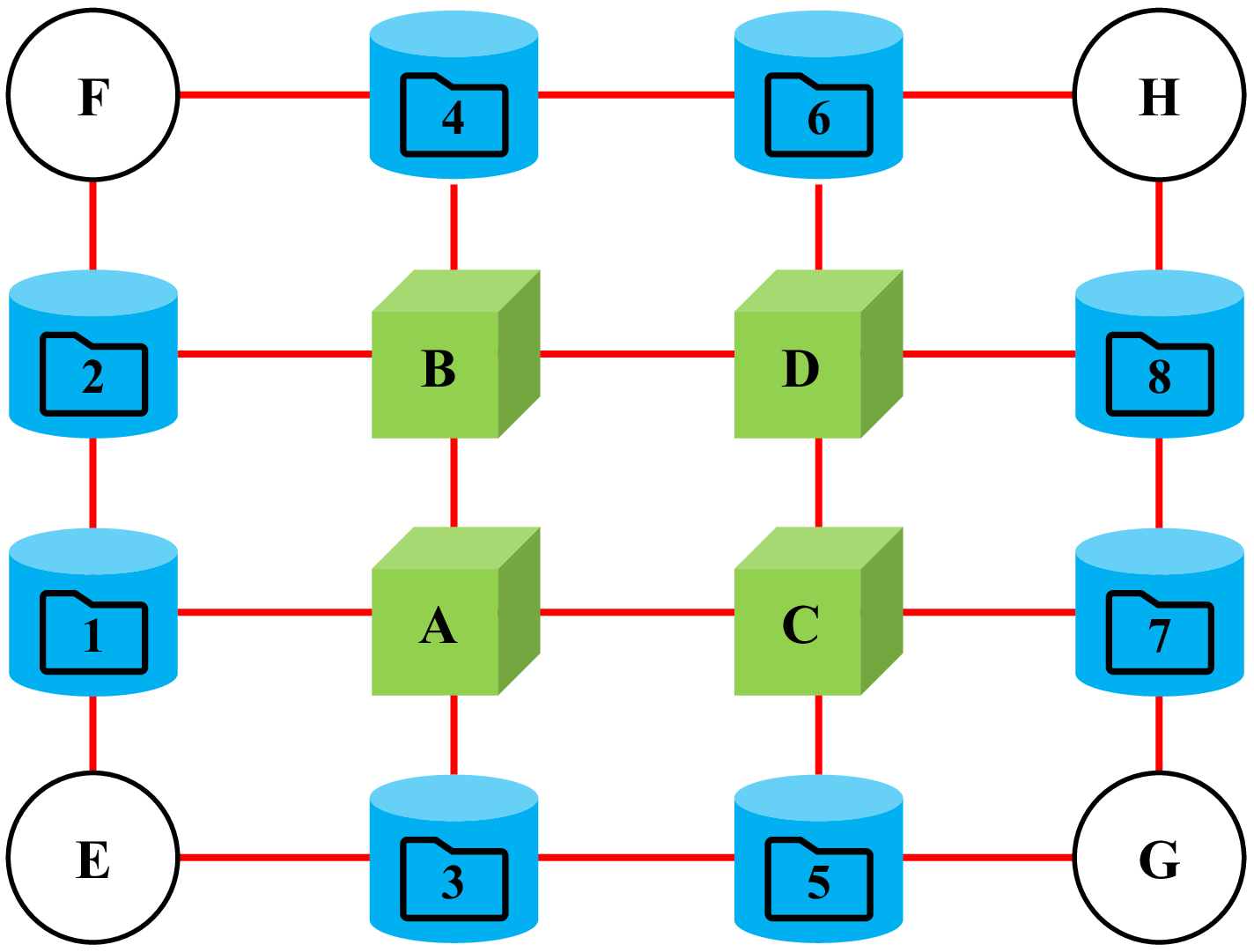}
        \caption{Network and stored databases.}
        \label{fig:network}
    \end{minipage}
    \hfill
    \begin{minipage}[t]{0.24\textwidth}
        \centering
        \includegraphics[width=.96\textwidth]{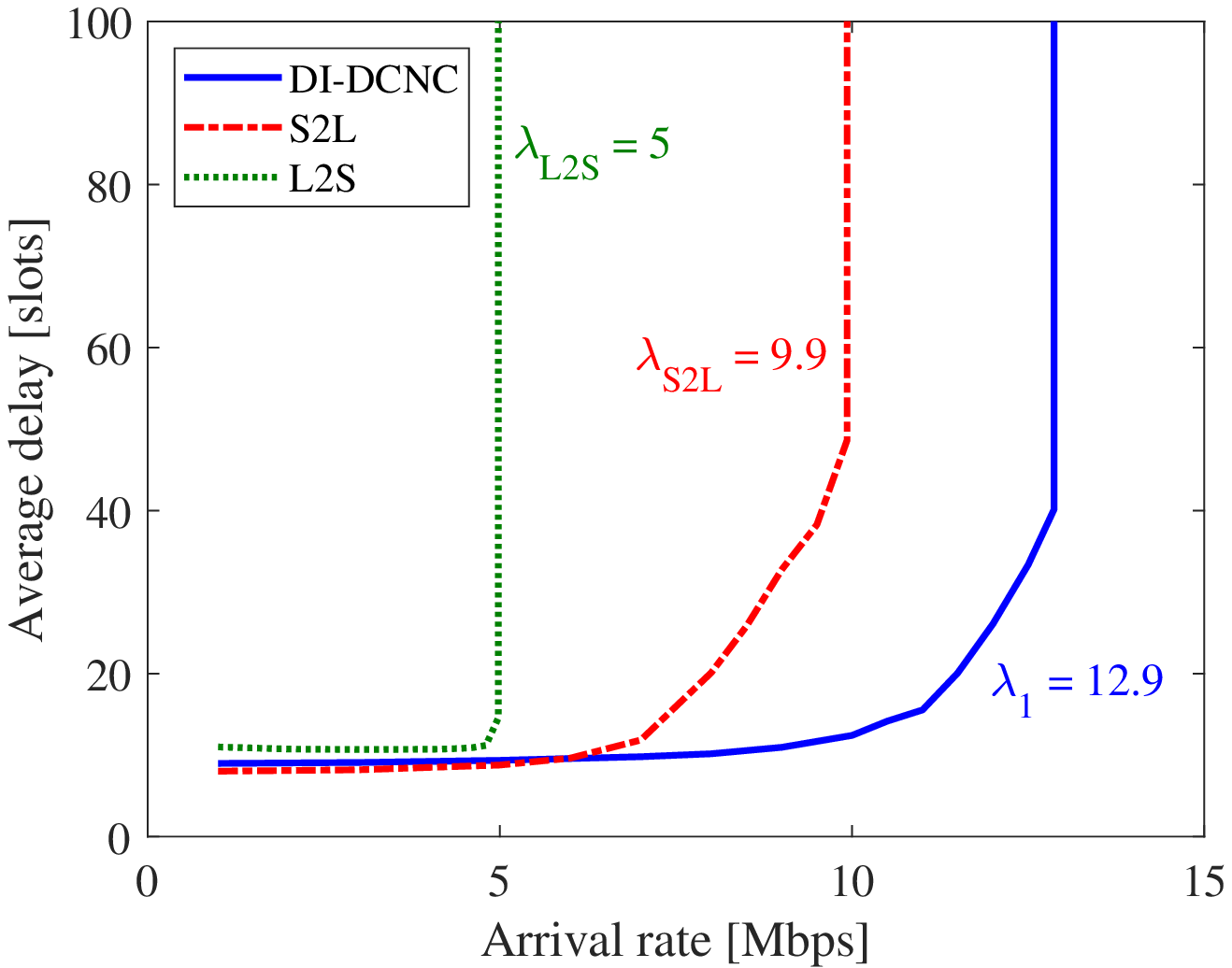}
        \caption{Network stability region.}
        \label{fig:network_stability_region}
    \end{minipage}
    \hfill
    \begin{minipage}[t]{0.24\textwidth}
        \centering
        \includegraphics[width=.96\textwidth]{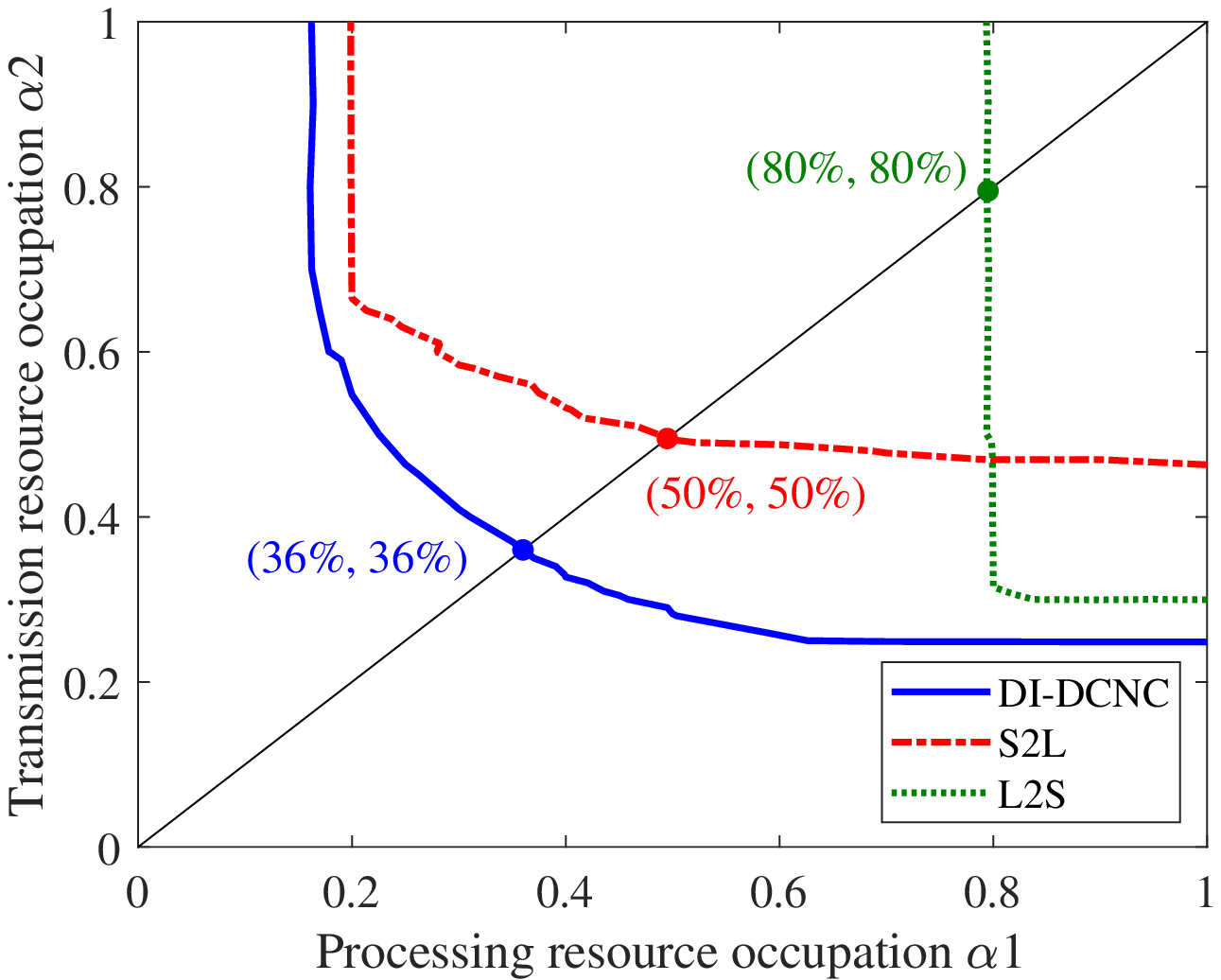}
        \caption{Resource occupation.}
        \label{fig:tradeoff}
    \end{minipage}
    \hfill
    \begin{minipage}[t]{0.24\textwidth}
        \centering
        \includegraphics[width=.96\textwidth]{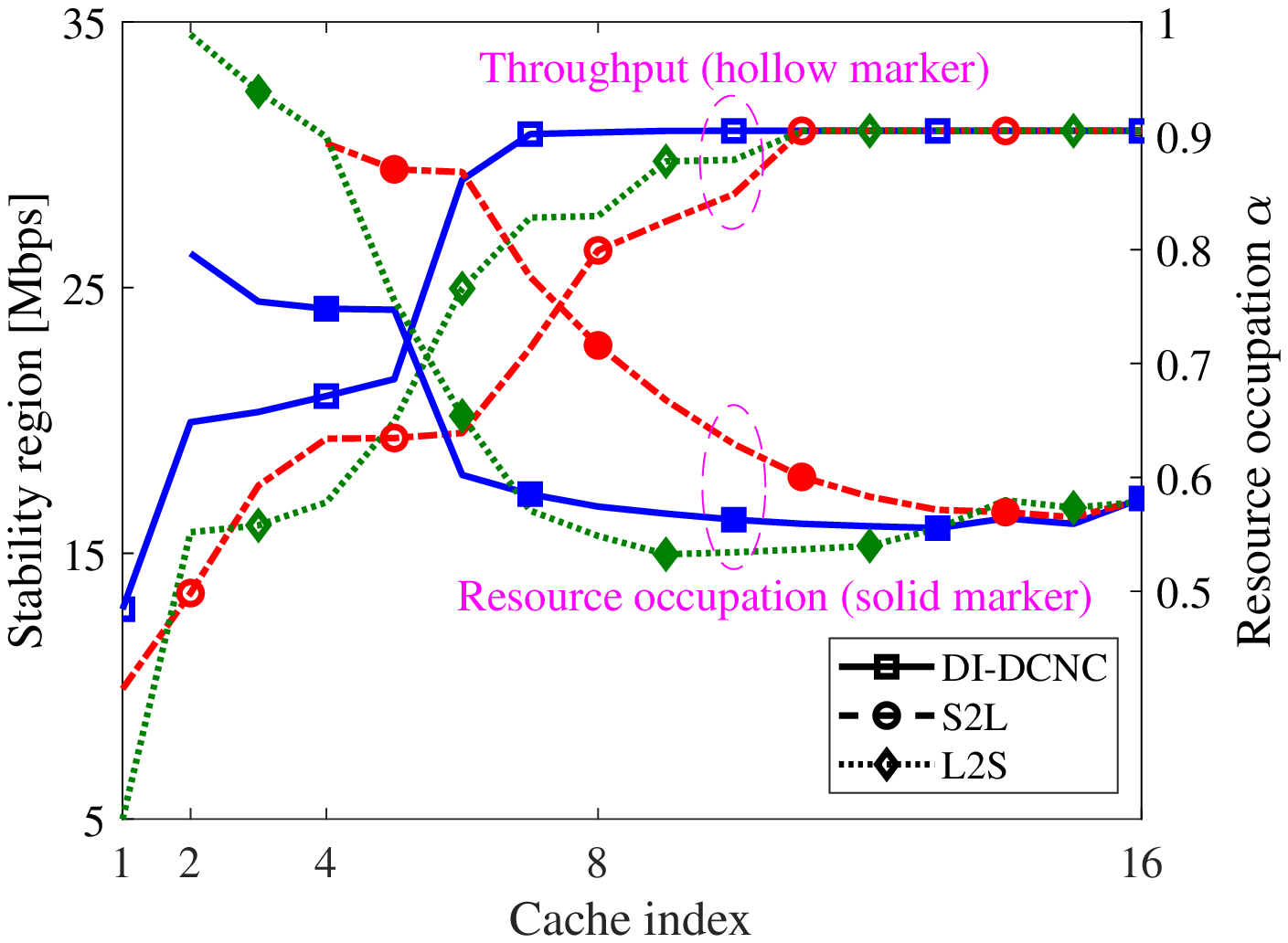}
        \caption{Effects of storage size.}
        \label{fig:cache_index}
    \end{minipage}
\end{figure*}

In this section, we evaluate DI-DCNC in the grid network in Fig. \ref{fig:network}, with the cached database at each node indicated by the index in the cylinder. The network resources are summarized in Table \ref{tab:resource}. We consider four clients, and each service includes $2$ functions, specified in Table \ref{tab:service} (to recall, client $=$ (source, destination, service), function $=$ (scaling factor, workload, object name, merging ratio)). The arrival processes are modeled by independent Poisson processes with $\lambda$ Mbps.

\begin{table}[ht]
    \centering
    \caption{Available Processing\,/\,Transmission Resources} \vspace{-5pt}
    \renewcommand\arraystretch{1.5}
    \begin{tabular}{l|l} \hline 
        Processing  & $C_i = 10$ GHz for $i = A, B, C, D$; $5$ GHz elsewhere \\ \hline 
        Transmission         &  $C_{ij} = 20$ Mbps for $\forall\, (i, j)$ \\ \hline 
    \end{tabular} \vspace{-15pt}
    \label{tab:resource}
\end{table}

\begin{table}[ht]
    \centering
    \caption{Clients $(s, d, \phi)$ and Function Spec $(\xi,\,r\, [\text{GHz}/\text{Mbps}], k, \zeta)$} \vspace{-5pt}
    \renewcommand\arraystretch{1.5}
    \begin{tabular}{l|c|c|c|c} \hline
        Client & $(E, H, \phi_1)$ & $(F, G, \phi_2)$ & $(G, F, \phi_3)$ & $(H, E, \phi_4)$ \\ \hline
        Func 1  & $(1, .2, 1, 1)$ & $(1, .5, 3, 2)$ & $(1, .1, 5, 1)$ & $(1/2, 1, 7, 5)$ \\ \hline
        Func 2  & $(2, .2, 2, 1)$ & $(1/2, .5, 4, 3)$ & $(3, .1, 6, 1)$ & $(1/3, 1, 8, 10)$ \\ \hline
    \end{tabular} \vspace{-5pt}
    \label{tab:service}
\end{table}

We employ two benchmark algorithms:
(i) Static-to-live (S2L), which makes individual routing decisions for live data, and then brings the static packets to the selected processing nodes.
(ii) Live-to-static (L2S), which brings live data to the ``nearest'' (in the weighted ALG) static source for processing.

\subsection{Network Stability Region}

First, we study the network stability regions of the algorithms, and Fig. \ref{fig:network_stability_region} depicts the attained average delay under different arrival rates (which is set to $\infty$ if the network is not stable).

As we can observe, the DI-DCNC algorithm attains good (average) delay performance over a wide range of arrival rates; when $\lambda$ crosses a {\em critical} point ($\lambda_1 \approx 12.9$ Mbps), the network is no longer stable, which is the boundary of its stability region. Similar behaviors are observed from the other two algorithms, and we find that DI-DCNC outperforms them in terms of the achieved throughput: $12.9$ Mbps (DI-DCNC) $>$ $9.9$ Mbps (S2L) $>$ $5.0$ Mbps (L2S). Therefore, compared to S2L and L2S, DI-DCNC can better exploit network resources to improve the throughput. We also notice that the delay attained by DI-DCNC is very similar and not lower than both benchmarks in the low-congestion regimes. This can be explained by the throughput-oriented design of DI-DCNC, which tries to reduce the {\em aggregate} queuing delay of both live and static data pipelines. Such objective, while closely related (especially in high congestion regimes), is not exactly equivalent to the actual service delay, which depends on the {\em maximum} delay between the two {\em concurrent} pipelines.

\subsection{Resource Occupation}

Next, we compare the resource occupation of the algorithms. Assume that the available processing/transmission capacities at each node/link are given by $\alpha_1$ and $\alpha_2$ (in {\em percentage}) of corresponding maximum budgets, respectively. We then define the {\em feasible region} as the collection of $(\alpha_1, \alpha_2)$ pairs under which the delay requirement is fulfilled. Assume $\lambda = 4$ Mbps and average delay constraint $= 20$ time slots.

The border lines of the feasible regions are shown in Fig. \ref{fig:tradeoff}. Since a lower latency can be attained with more resources, i.e., $(\alpha_1, \alpha_2) \to (1, 1)$, the feasible regions are to the upper-right of the border lines. As we can observe, DI-DCNC can save the most network resources. In particular, if $\alpha_1 = \alpha_2 = \alpha$, the resource saving ratios, i.e., $1 - \alpha$, of the algorithms are: $64\%$ (DI-DCNC) $>$ $50\%$ (S2L) $>$ $20\%$ (L2S). Another observation is: S2L is transmission-constrained (compared to its sensitivity to processing resource consumption). To wit: in the horizontal direction, it can achieve a maximal saving ratio of $80\%$ (when $\alpha_2 = 1$), which is comparable to DI-DCNC; however, the maximal ratio is $50\%$ for transmission resources (when $\alpha_1 = 1$), and there is a large gap to DI-DCNC in that dimension. The reason is that S2L ignores the transmission load of the static packets, leading to additional transmission resource consumption. While L2S is processing-constrained, because it is forced to use the processing resources at the static sources.

\subsection{Effects of Storage Resources}
\label{sec:caching}

Finally, we study the effects of available storage resources. We introduce the notion of {\em cache index}, which is the number of each database in the network.%
\footnote{
    In Fig. \ref{fig:network}, the cache index is $1$. If every node stores all databases, the cache index is $16$. We assume equal cache index for all databases.
}
We evaluate the same metrics of throughput and resource occupation, assuming $\lambda = 15$ Mbps, delay requirement $= 20$ slots, $\alpha_1 = \alpha_2 = \alpha$.

The results are shown in Fig. \ref{fig:cache_index}. For each algorithm, as the cache index grows, we observe increasing throughput and decreasing resource occupation,%
\footnote{
    In general, resource occupation reduces with cache index, but there are exceptions (see L2S curve in $[7, 13]$). A possible reason is that the developed algorithms do not explicitly minimize hop-distance, which can become more dominant than queuing delay in low congestion regimes (with low arrival rate or high cache index), making the end-to-end delay non-monotonic.
}
because with a higher cache index: for DI-DCNC and S2L, the distance to a static source reduces, while for L2S, processing resources at more network locations can be used. When comparing the three algorithms, we find that: (i) when the cache index hits $16$, the three algorithms become identical, and their performance converge. (ii) DI-DCNC achieves the best performance in most of the interval; in addition, it converges to the saturation point (the point after which the gain from increasing cache index becomes marginal) faster than other algorithms.

\section{Conclusions}

In this paper, we designed a cloud network control policy for online delivery of data-intensive \ac{agi} services. We considered a more general service model, requiring both live and static data-streams as inputs to the service function. Based on this model, we characterized the network stability region and proposed an efficient, throughput-optimal algorithm, DI-DCNC, which jointly decides (i) routing paths and processing locations for live data streams, and (ii) cache selection and distribution paths for associated data objects. Via numerical experiments, we demonstrated the superior performance of multiple pipeline coordination for the delivery of next-generation data-intensive real-time stream-processing services.

\ifCLASSOPTIONcaptionsoff
  \newpage
\fi

\bibliographystyle{IEEEtran}
\bibliography{IEEE_abrv,AgI}

\end{document}